\documentclass[amsmath,amssymb,prc,preprint,showpacs]{revtex4-2}

\usepackage{amssymb,amsmath} 
\usepackage{graphicx,epsfig} %for dealing with figures.

\newcommand {\mbf}[1]{{\mathbf{#1}}}

\begin{document}

\title{Coulomb screening in the momentum-space description of the  proton-deuteron elastic scattering: 
Why the renormalization is needed?}

%%%% To generate auto affiliation numbers please use \author{}\affil{} command

\author{A. Deltuva}
\email{arnoldas.deltuva@tfai.vu.lt}
\affiliation
{Institute of Theoretical Physics and Astronomy, 
Vilnius University, Saul\.etekio al. 3, LT-10257 Vilnius, Lithuania
}

\received{December 1, 2023}

\begin{abstract}%
  Proton-deuteron elastic scattering is considered in the framework of momentum-space Faddeev equations
  with the screening method for the Coulomb interaction.
  It is shown how the interplay of the proton-proton Coulomb potential and the deuteron pole in the neutron-proton
transition operator  leads to coinciding singularities in the Faddeev equation.
The coincidence of those singularities was not taken into account
 in a previous work
[Wita{\l}a et al., Eur.~Phys.~J.~A  41,  369  (2009)], 
leading to a conjecture that no renormalization is needed. However,
the coinciding singularities closely resemble those of the
point deuteron-proton system, naturally suggesting that 
the renormalization of the scattering amplitude is needed in the
unscreened Coulomb limit and can be performed conveniently in terms of the
point deuteron-proton system.
\end{abstract}

\maketitle

%\section{Introduction}
The proton-deuteron scattering in the momentum-space framework  is often 
described employing the screening of the Coulomb potential, thereby rendering it of short-range
and treatable by the standard scattering theory. However,
there is no consensus in the literature regarding  the behavior and renormalization of amplitudes 
in the unscreened limit. For example, while the most studies \cite{alt:78a,alt:80a,alt:02a,deltuva:05a}
require the renormalization of  scattering amplitudes,
the work by  Wita{\l}a et al. \cite{witala:09a} claims the existence of the unscreened limit for the
proton-deuteron elastic scattering amplitude without renormalization. 
The present work tries to explain those differences.

The long-range nature of the Coulomb interaction prevents the direct application of the standard
scattering theory. In the momentum space the Coulomb potential
\begin{equation}  \label{eq:w}
 \langle \mbf{p}_f| w | \mbf{p}_i \rangle \equiv w(\mbf{p}_f-\mbf{p}_i) =
 \frac{\alpha}{(\mbf{p}_f-\mbf{p}_i)^{2}}
 \end{equation}
is singular for the vanishing momentum transfer. In the
Lippmann-Schwinger equation for  the two-body transition operator 
\begin{equation}  \label{eq:t2b}
\langle \mbf{p}_f| t_c(p_o^2/2\mu+i0) | \mbf{p}_i \rangle = \frac{\alpha}{(\mbf{p}_f-\mbf{p}_i)^{2}}
%\langle \mbf{p}_f| w | \mbf{p}_i \rangle 
+ \int d^3p \, \frac{\alpha}{(\mbf{p}_f-\mbf{p})^{2}} \, \frac{2\mu}{p_o^2-p^2 + i0} \, 
\langle \mbf{p}| t_c(p_o^2/2\mu+i0) | \mbf{p}_i \rangle
\end{equation}
this potential singularity coincides with the singularity of the free resolvent, if the
 half-shell  condition $p_f = p_o$ is fulfilled. Although integrable separately, the coinciding
singularities become non-integrable.
 As a consequence, the equation is ill-defined and the standard scattering amplitude does not exist. 
 In the above equations  $\alpha = \alpha_{\rm e.m.}/2\pi^2$ with
 $\alpha_{\rm e.m.} \approx 1/137$ being the fine structure constant,
 $\mu$ is the reduced mass and $\mbf{p}_j$ denote 
the relative two-particle  momenta.

The problem can be solved by introducing the screening of the Coulomb potential at large distances $r > R$ in the 
coordinate space, rendering the momentum-space potential non-singular, then solving the Lippmann-Schwinger equation
and taking the unscreened Coulomb, i.e., $R \to \infty$ limit \cite{taylor:74a}. It was shown  that in the $R \to \infty$ limit
the half-shell and on-shell matrix elements of the transition operator acquire a diverging phase factor \cite{taylor:74a}.
After the isolation and removal of that factor, the so-called renormalization, the resulting amplitudes are 
well-behaved (at least, as distributions), and can be used to calculate scattering observables in a standard way. The procedure
can be extended when the additional short-range forces are present: the troublesome Coulomb contributions are isolated such that
the diverging phase factor to be removed via the renormalization is known from the pure Coulomb problem \cite{semon:75a}.

The idea of screening and renormalization has been applied also for the three-body problem;
see Refs. \cite{alt:78a,alt:80a,alt:02a,deltuva:05a} for a detailed description.
In short,
 full three-body transition operators are decomposed such that the problematic Coulomb contributions  are isolated.
In the proton-deuteron elastic scattering this is achieved by introducing auxiliary two-body operators driven
by the screened Coulomb force between the proton and the center-of-mass of the deuteron. Being solutions of the 
two-body Coulomb problem, these operators display the corresponding behavior in the $R \to \infty$ limit
where the diverging phase factor is known and can be removed via the renormalization.
It must be emphasized that these  auxiliary operators do not imply approximating the proton-proton
Coulomb  force by the proton-deuteron one. In the solved three-body equations
the Coulomb force between the protons is  included.
Wita{\l}a et al. \cite{witala:09a} have chosen a slightly different strategy. They started with the symmetrized three-nucleon
Faddeev equation in the isospin formalism, where the two-body proton-proton transition operators contained also
Coulomb contributions. Without introducing auxiliary operators, they investigated the behavior of all terms
in the Faddeev equation and concluded the absence of strong singularities for the elastic amplitude.
This led to the conclusion that the proton-deuteron elastic scattering amplitude should exist without renormalization. 

In contrary to Ref.~\cite{witala:09a},
 it will be shown here that also without the introduction of the auxiliary operators
the proton-deuteron elastic scattering equation in the unscreened Coulomb limit has coinciding singularities
of the same type as in the two-body problem (\ref{eq:t2b}). There is no intention to provide
a complete re-derivation of the proton-deuteron formalism which is already given
in  Refs. \cite{alt:78a,alt:80a,alt:02a,deltuva:05a}; the Faddeev equation of Ref.~\cite{witala:09a} is
not the optimal starting point for this purpose.

In the following, the dependence on the Coulomb screening radius $R$ is suppressed in the notation, having
the unscreened limit $R \to \infty$ in mind.
The Faddeev equation in the
Alt-Grassberger-Sandhas (AGS) version \cite{alt:67a}
for the symmetrized proton-deuteron transition operator
\begin{equation} \label{ags}
U =  PG_0^{-1} + PtG_0U 
\end{equation}
after one iteration becomes
\begin{equation} \label{ags-it}
U =  PG_0^{-1} + PtP + PtG_0 PtG_0 U .
\end{equation}
Here  $G_0$ is the free resolvent, $t$ is the two-nucleon transition operator, and  $P$ is
the sum of two cyclic permutation  operators \cite{deltuva:05a}.
Isospin formalism is used, meaning that for given isospin states
the $t$-operator is a linear combination of  proton-neutron and proton-proton  operators,
the latter including also the Coulomb contribution.
For calculation of elastic scattering observables one needs the matrix elements of $U$ between 
the proton-deuteron states $| \phi_d \mbf{q}_j \rangle$ where
$\phi_d$ denotes the deuteron wave function with the binding energy  $\epsilon_d$,
and $\mbf{q}_j$ denotes the relative spectator-pair momentum. The dependence on the discrete
quantum numbers is suppressed in the notation.
With explicit  intermediate integrations  the third term in Eq. (\ref{ags-it})
then reads
\begin{equation} \label{eq:ptpt}
 \int \langle \phi_d \mbf{q}_f |PtP|\mbf{p} \mbf{q}\rangle d^3p \, d^3q  \langle \mbf{p} \mbf{q} |
  G_0 tG_0 | \mbf{p}' \mbf{q}'\rangle d^3p'\, d^3q' \langle \mbf{p}' \mbf{q}' |  U | \phi_d \mbf{q}_i \rangle .
\end{equation}
The structure of this integrand will be investigated in the
proton-deuteron on-shell limit for $q$, that is,
$q \to q_o$, where $q_o$ is the relative proton-deuteron on-shell momentum, corresponding to 
the system energy in the c.m. frame $E=q_o^2/2\mu_{pd} - \epsilon_d$. Near this $q$ value 
 the matrix elements of the  free resolvent  $G_0$ are finite, but
 the two-nucleon transition operator has a pole corresponding to the deuteron bound state, i.e.,
\begin{equation} \label{dpole}
  \langle \mbf{p} \mbf{q} |
  G_0 tG_0 | \mbf{p}' \mbf{q}'\rangle  \xrightarrow[q \to q_o]{}
  \langle \mbf{p} | \phi_d \rangle \frac{2\mu_{pd} \, \delta(\mbf{q} - \mbf{q}')}{q_o^2 - q^2 + i0}
  \langle \phi_d |\mbf{p}'\rangle.
\end{equation}
Thus,  in the vicinity of $q \to q_o$ the
contribution of the deuteron pole to (\ref{eq:ptpt}) is
\begin{equation}
  \int \langle \phi_d \mbf{q}_f |PtP|\phi_d \mbf{q}\rangle  d^3q \, \frac{2\mu_{pd}}{q_o^2 - q^2 + i0}
  \langle \phi_d \mbf{q} |  U | \phi_d \mbf{q}_i \rangle .
\end{equation}

The matrix element $\langle \phi_d \mbf{q}_f |PtP|\phi_d \mbf{q}\rangle$,
appearing also as the second term in Eq. (\ref{ags-it}),
in the isospin formalism
has contributions from proton-proton and neutron-proton $t$-operators, which are fully off-shell 
under the $q \to q_o$ condition. The non-singular contributions are omitted in the following. 
In the unscreened Coulomb limit $R \to \infty$ the singular part of this matrix element
arises from the off-shell proton-proton transition operator $t_c^R$
contained in $t$
whose singularity is the same as 
in the proton-proton Coulomb potential \cite{kok:80a}. Thus, when investigating the structure of singularities
one can  replace $t$ by $\frac23 \, w$, where the $\frac23$ factor arises from the isospin
weighting \cite{deltuva:05a}. The result has the form
\begin{equation} \label{eq:pwp}
  \frac23 \langle \phi_d \mbf{q}_f |P w P|\phi_d \mbf{q}\rangle
=  w(\mbf{q}_f - \mbf{q}) \, F_1(\mbf{q}_f - \mbf{q}) +  F_2(\mbf{q}_f,\mbf{q})
\end{equation}
where the regular form-factor type functions $F_k$ involve spin-coupling coefficients
and integrals over the deuteron wave function ($F_2$ involves also the Coulomb potential, but its singularity is integrated out).
Detailed expressions for  $F_k$ are irrelevant, except for the particular feature
$F_1(0)=1$. This means that in the limit of the  vanishing momentum transfer $\mbf{q}_f \to \mbf{q}$,
%i.e., for the proton-deuteron forward scattering,
 the singular part
of (\ref{eq:pwp}) is exactly the same as in the proton scattering off a point deuteron.
Combining together expressions (\ref{dpole}) and (\ref{eq:pwp}) in the unscreened Coulomb limit
and leaving out regular terms one arrives at 
\begin{equation} \label{eq:sing}
  \int d^3q \, \frac{F_1(\mbf{q}_f - \mbf{q})}{(\mbf{q}_f - \mbf{q})^2} \,
  \frac{2\mu_{pd}}{q_o^2 - q^2 + i0} \,
  \langle \phi_d \mbf{q} |  U | \phi_d \mbf{q}_i \rangle .
\end{equation} 
Obviously, in the proton-deuteron on-shell limit $q_f=q_o$ the two singularities in (\ref{eq:sing}) coincide,
the integral does not converge. Furthermore,
the structure of singularities is the same as in the two-body pure Coulomb problem (\ref{eq:t2b}).
One may also note that second term of (\ref{ags-it}) at $q_f=q_o=q_i$ and  $\mbf{q}_f \to \mbf{q}_i$ 
 would include an infinite contribution
(\ref{eq:pwp}) to the forward scattering amplitude, resembling the proton-deuteron Coulomb potential.
Thus, there is a close similarity with the two-body (proton + point deuteron) scattering problem
involving short-range and Coulomb forces, and one may expect similar mathematical properties when applying the method
of Coulomb screening. In particular,  the 
proton-deuteron elastic scattering amplitudes would acquire diverging
phase factors in the $R \to \infty$ limit and would necessitate  the renormalization.
Based on the above equations, one could expect the proton scattering off a point deuteron to be the reference
problem that determines renormalization factors and direct Coulomb amplitude, as the effective charge, momenta, and 
reduced mass correspond exactly to those of the (proton + point deuteron) system.
Noteworthy, this conjecture is achieved without introducing auxiliary proton-deuteron operators. Thus, the renormalization in terms of the
(proton + point deuteron) system is not introduced artificially but
emerges naturally from the analysis of singularities.

As the above derivation was not relying on the properties of the initial state $| \phi_d \mbf{q}_i \rangle $,
the singularity structure made explicit in (\ref{eq:sing}) obviously persists for any initial momentum $\mbf{q}_i$
and the pair state with relative momentum  $\mbf{p}_i$.
In analogy with the two-body problem this again implies that not only the on-shell but 
also  the half-shell matrix elements  $\langle \phi_d \mbf{q}_f |  U | \mbf{p}_i \mbf{q}_i \rangle$ at
$q_f=q_o \neq q_i$ need an appropriate renormalization. The initial-final state symmetry of the transition operator $U$ suggests 
the necessity of the renormalization also at  $q_i = q_o \neq q_f$. The singularity structure at $q_i = q_o$
can be easily shown to be of the type (\ref{eq:sing}). For this one could
 start with an alternative form of the AGS equation
\begin{equation}
U =  PG_0^{-1} + UG_0tP 
\end{equation}
and repeat similar steps as given in (\ref{ags-it}) - (\ref{eq:pwp})
for the half-shell matrix element $\langle \mbf{p}_f  \mbf{q}_f |  U | \phi_d \mbf{q}_i \rangle$,
 arriving at
\begin{equation} \label{eq:singb}
  \int d^3q \, \langle  \mbf{p}_f \mbf{q}_f |  U | \phi_d \mbf{q} \rangle \, 
  \frac{2\mu_{pd}}{q_o^2 - q^2 + i0} \,
\frac{F_1(\mbf{q} - \mbf{q}_i)}{(\mbf{q} - \mbf{q}_i)^2} 
\end{equation} 
instead of (\ref{eq:sing}), with coinciding singularities at $q_i=q_o$.

The fact that half-shell matrix elements $\langle \mbf{p}_f  \mbf{q}_f |  U | \phi_d \mbf{q}_i \rangle$ at
$q_i=q_o$ need an appropriate renormalization implies a corresponding property also for operator products 
acting on $ U | \phi_d \mbf{q}_i \rangle$. In particular, 
the Faddeev operator of  Ref.~\cite{witala:09a}
\begin{equation} \label{eq:wt}
tG_0U \equiv T =  tP + tG_0PT,
\end{equation}
when acting on the initial proton-deuteron state with $q_i=q_o$, i.e.,
$ T | \phi_d \mbf{q}_i \rangle = t G_0 U | \phi_d \mbf{q}_i \rangle $,
at least  needs a  renormalization associated with the initial state. 
The elastic amplitude \cite{witala:09a} is
\begin{equation} \label{eq:wtel}
\langle \phi_d \mbf{q}_f |(P G_0^{-1} +  P T )| \phi_d \mbf{q}_i \rangle =
\langle \phi_d \mbf{q}_f |P G_0^{-1} | \phi_d \mbf{q}_i \rangle +
\langle \phi_d \mbf{q}_f |  P t P | \phi_d \mbf{q}_i \rangle +
\langle \phi_d \mbf{q}_f |  P t P G_0 T | \phi_d \mbf{q}_i \rangle
\end{equation}
where one can easily identify the three terms in (\ref{ags-it}),
a consequence of the equivalence between the AGS equation (\ref{ags}) for $U$ and the Faddeev equation (\ref{eq:wt}) for $T$.
 In particular,
since in the two-nucleon ${}^3S_1-{}^3D_1$ wave the  operator
$T$ carries the deuteron bound-state pole as $t$ does, the singularity structure of the third term in 
(\ref{eq:wtel}) is exactly the same as in (\ref{eq:sing}), i.e., the two singularities from the proton-proton
Coulomb potential and the deuteron pole do coincide, such that the renormalization in the unscreened limit of 
(\ref{eq:wtel}) is needed also for the final state. 
On the other hand,  looking back to the left-hand sides of
 (\ref{eq:wt}) and (\ref{eq:wtel})
 it is quite obvious that $T$ must exhibit a singular behavior itself,
stemming from the off-shell $t_c^R$.

The above consideration is not a rigorous proof of the screening and renormalization procedure for
the proton-deuteron scattering, see Refs.~\cite{alt:80a,deltuva:05a} instead. However, it demonstrates one more time that
in the proton-deuteron scattering problem one encounters typical difficulties related to the Coulomb treatment, manifesting themselves
by coinciding singularities in the momentum-space framework. This feature arises not from  the proton-proton 
transition operator directly, but via the interplay between the proton-proton Coulomb interaction and
the deuteron bound-state pole. In fact, one may consider the energy regime
below the deuteron breakup threshold where the two-nucleon transition operators are always off-shell at negative two-nucleon
energies, and do not need renormalization. However, even in this case the coinciding singularities (\ref{eq:sing}) persist
and, consequently, the proton-deuteron elastic scattering amplitudes need renormalization.

It is very important to note
that the work by Wita{\l}a et al. \cite{witala:09a} disregarded
the appearance of coinciding singularities by interplay of the proton-proton Coulomb  potential and the deuteron pole.
 In section 4 of Ref.~\cite{witala:09a} Wita{\l}a et al.
considered only the singularities and renormalization features of  the half-shell and on-shell proton-proton Coulomb transition matrix
and the free resolvent, that indeed do not coincide in the expression for the elastic amplitude.
However, they missed the fact that the off-shell proton-proton transition matrix, though exists in the unscreened
limit without renormalization, nevertheless carries the singularity of the Coulomb potential, which
together with the deuteron pole leads to coinciding singularities in proton-deuteron scattering equations.
For example, the analysis of the off-shell $t_c^R$
in the unscreened limit of equation (D.8) of  Ref.~\cite{witala:09a} reveals that this takes place for
$|\mbf{p} \pm \mbf{q}/2| = q_0$, in combination with $P$  leading to the appearance of coinciding singularities of type
(\ref{eq:sing}) in (D.9), that are not taken into account in the investigation of the unscreened limit.
This way Wita{\l}a et al.
 arrived at an erroneous conclusion that the proton-deuteron elastic scattering amplitude in
 the unscreened limit should exist without renormalization. 
Noteworthy, the numerical studies of  Ref.~\cite{witala:09a} due to technical reasons
omitted the term 
$\langle \phi_d \mbf{q}_f |  P t_c^R P G_0 T | \phi_d \mbf{q}_i \rangle$,
which in the present work is taken as an example for the appearance of
 coinciding singularities in the $R \to \infty$ limit.
The remaining terms in the elastic amplitude of Ref.~\cite{witala:09a}  cancel partially,
possibly reducing the sensitivity of the phase at finite $R$.
The shortcoming of  Ref.~\cite{witala:09a}  persists also in subsequent studies of the breakup reaction,
and also in a recent preprint \cite{witala:24p}. The latter work mostly deals with the decomposition into contributions of low partial waves 
and the remaining three-dimensional (3D) ones and partial cancellations between them.
Considerations  of the present work are performed directly in 3D,
which, although not used in practical calculations \cite{deltuva:05a,deltuva:09e}, is more transparent for the analysis of
singularities.
 Finally, the numerical studies \cite{deltuva:09e,adc-bonn} using full amplitudes
clearly confirmed the need for the  renormalization of the scattering amplitudes
consistently with Refs.~\cite{alt:80a,deltuva:05a}.

\vspace{1mm}

\bibliographystyle{prsty}
%\bibliography{abbrev,hann,book,pre80,80-89,90-99,200x,clmb,ad,adconf,exp}

 \end{document}